\documentclass[a4paper]{jpconf}
\usepackage{graphicx}
\begin{document}
\title{Brane-world black holes}

\author{Panagiota Kanti}

\address{Division of Theoretical Physics, Department of Physics, University of Ioannina,
Ioannina GR 45110, Greece}

\ead{pkanti@cc.uoi.gr}

\begin{abstract}
In this talk, I present and discuss a number of attempts to construct black hole
solutions in models with Warped Extra Dimensions. Then, a contact is made with models
with Large Extra Dimensions, where black-hole solutions are easily constructed -- here
the focus will be on the properties of microscopic black holes and the possibility of
using phenomena associated with them, such as the emission of Hawking radiation,
to discover fundamental properties of our spacetime.
\end{abstract}

\section{Introduction}

A breakthrough in string theory \cite{kanAntoniadis, kanHW, kanLykken}, that decoupled
its fundamental scale from the Planck scale, led to the construction of two
higher-dimensional gravitational models a decade ago: the scenario with Large Extra
Dimensions \cite{kanADD, kanAADD} and the one with Warped Extra Dimensions
\cite{kanRS} (for some early higher-dimensional gravitational models, see 
\cite{kanAkama, kanRuS, kanVisser, kanWiltshire}). Both models use the idea
of a 3-brane that plays the role of our 4-dimensional world and is embedded in
a higher-dimensional bulk: $n$ flat, spacelike dimensions in the first case,
and a fifth spacelike dimension that is part of an Anti de Sitter spacetime
in the second. In both cases, all ordinary matter is localised on the brane, while
gravitons have access to the whole spacetime.  

The main point in both models is that the introduction of a higher-dimensional
spacetime is supplemented by a new fundamental scale. In the Large Extra Dimensions
scenario the higher-dimensional theory has a new scale for gravity, $M_{*}$,
that is related to the effective 4-dimensional Planck scale through the equation 
$M_P^2 \simeq {\cal R}^{\,n} \,M_{*}^{2+n}$ \cite{kanADD, kanAADD}. If the size
of the extra dimensions satisfies the inequality ${\cal R} \gg l_P$, $M_*$ can
be significantly lower than the 4-dimensional one $M_P$, and the fundamental
gravitational constant
$G_D=1/M_*^{n+2}$ much larger than the 4-dimensional one $G_D$. In the case of
Warped Extra Dimensions, it is the Electroweak scale that is derived from the
fundamental one through the relation $M_{EW}=e^{-kL}\,M_*$, where $L$ is the
distance between our observable brane and a hidden one, and $k$ the curvature
scale associated with the negative cosmological constant that fills the 5-dimensional
spacetime of the model \cite{kanRS}. 

In the context of these two scenaria, we are looking for black hole solutions. 
As we will see, the low scale for gravity of the Large Extra Dimensions will have
important consequences for the creation and evaporation process of black holes in these
theories. On the other hand, the richer topological structure of the model with
Warped Extra Dimensions will prove to be a major obstacle for the construction 
of black-hole solutions in these theories. We will thus start with the latter class
of models and present a number of approaches to this problem, their results, and 
the new questions emerging. We will see how a natural contact is made between the
two scenaria when it comes to black holes, and thus we will move to the Large Extra
Dimensions scenario for the study of the much simpler, and easier to construct,
black-hole solutions. A brief study of the creation process of the microscopic
black holes and their properties will be made, and many current results on the
characteristics of their evaporation process will be presented (for a number of
reviews where more information is offered on black-hole solutions in both types
of models, see \cite{kanreview1, kanreview, kanreview2}). 

%%%%%%%%%%%%%%%%%%%%%%%%%%%%%%%%%%%%%%%%%%%%%%%%%%%%%%%%%%%%%%%%%%%%%%%%%%%%%%%%

\section{The Warped Case}

Despite the presence of additional spacelike dimensions in the Large and Warped
Extra Dimensions scenaria, our conception of the creation of a black hole has not
changed: whenever ordinary matter, which according to the assumptions of the models
is localised on our brane, undergoes gravitational collapse, we expect a black hole
to form. We also anticipate that the produced black holes will have the following
characteristics:  
\begin{itemize}
\item[-] {\it Location\,:} Being a gravitational object, the black hole will be,
at least initially, attached to our brane but it will also extend off our brane

\item[-] {\it Geometry\,:} Due to the warping along the fifth dimension in the
Warped Extra Dimensions Scenario, we expect the black hole to have an additional
cylindrical symmetry

\item[-] {\it Asymptotics\,:} Due to the localization of the source on the brane,
the black hole should have a regular asymptotic form both on the brane (Minkowski, de Sitter
or Anti de Sitter) and off the brane (Anti de Sitter).
\end{itemize}
%%%%%%%%%%%%

However, in practice, the task of constructing a black hole solution with the
above characteristics has proven to be extremely challenging. The attempts to overcome this
difficulty have led to different approaches of the problem, some of which are discussed
below:

\subsection{{\it ``The-brane-observer-comes-first'' point of view}} Starting from
a pure 4-dimensional Schwarzschild solution on the brane, the authors of \cite{kanCHR}
embedded it in a 5-dimensional Anti de Sitter spacetime. The line-element then 
assumed the form:
%%%%%%%%%%%%%%
\begin{equation}
ds^2=e^{-2 k |y|} \left[-\left(1-\frac{2M}{r}\right) dt^2 +
\left(1-\frac{2M}{r}\right)^{-1} dr^2 + r^2\,d\Omega_2^2\right] + dy^2\,.
\label{kanCHRmetric}
\end{equation}
%%%%%%%%%%%%%%
The above ansatz was an attractive choice as it satisfied the Randall-Sundrum field
equations without the need for any extra bulk matter $T_{MN}$ -- since the Schwarzschild 
solution is a vacuum solution and the above line-element was of a factorised type --
and its projection on the brane was a well-known black hole solution. However, the
five-dimensional background was {\it not}. By using the above line-element, one
easily finds that
%%%%%%%%%%%
\begin{equation}
R_{MNRS}\,R^{MNRS} \sim \frac{48 M^2 e^{4 k |y|}}{r^6}\,.
\label{kanInvCHR}
\end{equation}
%%%%%%%%%%%%
The above results reveals that the metric (\ref{kanCHRmetric}) describes a black string
and not a black hole as the singularity at $r=0$ extends over the whole range of the
fifth coordinate. As it was later shown \cite{kanRuthGL}, the same solution was plagued
by the well-known Gregory-Laflamme \cite{kanGL} instability of black string solutions that
turns the black string into an infinitely long row of `black cigars'.

\subsection{{\it ``The-brane-observer-is-all-that-matters'' point of view}} In this
approach \cite{kanMG1}, the view was taken that although it is indeed difficult to find an
analytic solution describing a 5-dimensional localised-close-to-the-brane black hole, one
may still study its 4-dimensional projection on our brane, which is, in any case, what a
brane observer would ever see. For the construction of the 4-dimensional projected
line-element, the following brane field equations were used \cite{kanMST}
%%%%%%%%%%%%%%%%%%
\begin{equation}
^{(4)}G_{\mu\nu}=8\pi G_4 T_{\mu\nu} + (8\pi G_5)^2\,\pi_{\mu\nu} + E_{\mu\nu}\,,
\label{kanBrEq}
\end{equation}
%%%%%%%%%%%%%%%%%
where $\pi_{\mu\nu}$ is a tensor quadratic to the energy-momentum tensor $T_{\mu\nu}$
-- and thus may be ignored in the low-energy limit -- and $E_{\mu\nu} = ^{(5)}C_{y\mu y \nu}$
is a projected-on-the-brane 5-dimensional Weyl tensor. Although brane black hole solutions
could be derived from the above field equations, this could only be done by employing
certain assumptions about the form of $E_{\mu\nu}$, whose exact expression can follow
only by solving the complete 5-dimensional gravitational problem. 

\subsection{{\it ``Restore-the-bulk'' point of view}} 
\begin{figure}[t]
\begin{center}
\includegraphics[width=7.5cm, height=6.0cm]{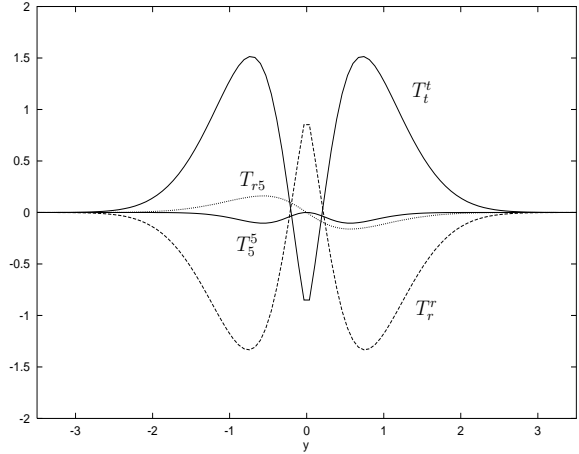}
\end{center}
\caption{\label{kanshellKT} The profile of the bulk energy-momentum tensor $T_{MN}$ along
the fifth dimension \cite{kanKOT}.}
\end{figure}

Returning to the approach taken
by the authors of \cite{kanCHR}, an alternative idea was proposed in \cite{kanKOT} 
according to which the 5-dimensional line-element assumed the non-factorised form
%%%%%%%%%%%%%
\begin{equation}
ds^2=e^{-2 k |y|}\left[-\left(1-\frac{w(y)}{r}\right)dt^2 +
\left(1-\frac{w(y)}{r}\right)^{-1} dr^2 + r^2\,d\Omega_2^2\right] + dy^2\,.
\label{kanKOTmetric}
\end{equation}
%%%%%%%%%%%%%%%
For the above line-element to satisfy the 5-dimensional bulk equations, an additional
bulk energy momentum tensor had to be introduced. Its presence then guaranteed that
a localising mechanism for the black-hole singularity was in action if we demanded
that 
%%%%%%%%%%%%%%%
\begin{equation}
w(y) \rightarrow 2M \quad (y=0)\,, \qquad w(y) \rightarrow 0 \quad (y\rightarrow \infty)\,.
\end{equation}
%%%%%%%%%%%%%%
When the bulk equations were solved for the expression of the energy-momentum tensor, the
profile, shown in Fig. \ref{kanshellKT}, was found. The bulk matter had a shell-like
distribution along the extra dimension thus engulfing the brane and restricting the
extent of the black hole singularity. In addition, it satisfied a stiff equation of
state with $p_r \simeq \rho$, and it obeyed the energy conditions on the brane: $\rho>0$
and $\rho+p_i>0$. The main drawback of the model was that conventional scalar or gauge
field configurations failed to support such an energy-momentum tensor.

%%%%%%%%%%%%%%%%%%%%
%%%%%%%%%%%%%%%%%%%%%%

\subsection{{\it ``The-bulk-comes-first'' point of view}}
\begin{figure}[t]
\begin{center}
\includegraphics[width=7cm, height=9.0cm]{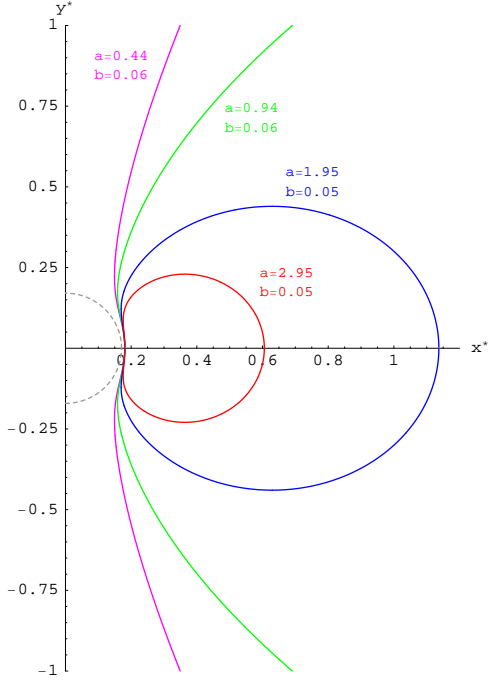}
\end{center}
\caption{\label{kanCGKMplot} Brane trajectories in a 5D Schwarzschild-AdS background \cite{kanCGKM}.}
\end{figure}

A different approach was taken in \cite{kanCGKM} where a regular 5-dimensional
black hole solution was put at `the center of the stage' (see also \cite{kanGGI}).
The line-element was assumed to be of the form
%%%%%%%%%%%%%%%%%
\begin{equation}
ds^2=-U(r)\,dt^2 + \frac{dr^2}{U(r)} + r^2\,(d\chi^2 +\sin^2\chi\,d\Omega_2^2)\,,
\label{kanCGKMmetric}
\end{equation}
%%%%%%%%%%%%%%%%%
describing a black hole in Anti de Sitter spacetime with
%%%%%%%%%%%%%%%%
\begin{equation}
U(r)=1+k^2r^2-\frac{\mu}{r^2}\,.
\end{equation}
%%%%%%%%%%%%%%
In this background, a 3-brane was then introduced through the junction conditions
$[K_{\mu\nu}-K\,h_{\mu\nu}]^+_-=\kappa_5\,T_{\mu\nu}$ \cite{kanIsrael},  where $h_{\mu\nu}$
is the induced metric on the brane, $K_{\mu\nu}$ the extrinsic curvature and $T_{\mu\nu}$
the brane energy-momentum tensor of a perfect fluid. Since $U(r)$ was known, the
junction conditions formed a system of coupled differential equations for $\chi(t,r)$
(the trajectory of the brane), $\rho(t,r)$ (the brane energy density) and $w(t,r)$
(the brane equation of state). After a detailed analytical study, the following
conclusions were drawn: (i) all regular trajectories stayed clear of the bulk
black-hole horizon, (ii) no regular black-hole solution with the singularity on
the brane was found, and (iii) the most physically interesting solution had $\rho$
and $p$ on the brane resembling those of a star (for a set of brane trajectories,
see Fig. \ref{kanCGKMplot}). 
 
%%%%%%%%%%%%%%%%%%%%
%%%%%%%%%%%%%%%%%%%%%%

\subsection{{\it ``Should-these-BH's-exist-at-all?'' point of view}}

By using the AdS/CFT correspondence, in \cite{kanTanaka} the action of the Randall-Sundrum
model was written as 
%%%%%%%%%%%%%
\begin{equation}
S_{RS}=S^{(4)}_{EH} + 2 W_{CFT}\,,
\end{equation}
where $S^{(4)}_{EH}$ is the Einstein-Hilbert action and $W_{CFT}$ the connected
Green function of a Conformal Field Theory in 4 dimensions. According to the above,
in the context of the 4-dimensional CFT, if a black hole is formed, it does so in the
presence of a large number of conformal fields. This will inevitably induce emission
of Hawking radiation of CFT modes with a significant back-reaction to the metric. 
As a result, a static localised black hole with $r_H \gg \ell_{AdS}$ (the regime
in which the AdS/CFT correspondence is valid) may {\it not exist} at all in a
brane-world theory. Similar claims were made around the same time in the context
of other works (see, for example, \cite{kanBGM, kanEFK}).

With the difficulty in constructing analytically regular black hole solutions in
warped models having been clearly established, numerical analyses were undertook next.
In \cite{kanKTN} the following ansatz for the 5-dimensional black-hole line-element was
considered 
%%%%%%%%%%%%%%%
\begin{equation}
ds^2=\frac{\ell_{AdS}^2}{z^2}\,\left[-T^2 dt^2 +
e^{2R}\,(d\rho^2+\frac{\rho^2}{4 \xi}\,d\xi^2) + r^2\,e^{2C}\,d\Omega_2^2\right]\,.
\end{equation}
%%%%%%%%%%%%%%%%%
The boundary conditions supplementing the desired solution were: (i) the existence of
a horizon at a constant value $\rho_h$, (ii) the demand that the metric functions and
their derivatives are finite at $\rho_h$, and (iii) that at asymptotic infinity, the
Anti de Sitter spacetime is recovered with $T \rightarrow 1$ and $R,C \rightarrow 0$.
In the numerical analysis that was performed, small, static, asymptotically AdS 
black holes were indeed found but with $\rho_h \le \ell_{AdS}/5$. A similar analysis
in 6 dimensions \cite{kanKudoh} extended that limit up to $\rho_h =2 \ell_{AdS}$. No
larger black holes in a warped spacetime have been found numerically so far, a result
that seems to support the claims of \cite{kanTanaka, kanBGM, kanEFK}.

A recent numerical analysis performed in \cite{kanTT} followed the construction method
of \cite{kanCGKM} and considered a 5-dimensional topological black hole in Anti de Sitter
spacetime with line-element
%%%%%%%%%%%%%%%
\begin{equation}
ds^2=-U(r)\,dt^2 + \frac{dr^2}{U(r)} + r^2\,d\Omega_{3(\beta)}^2\,,
\end{equation}
with
%%%%%%%%%%%%%%
\begin{equation}
U(r)=\beta+k^2r^2-\frac{\mu}{r^2}\,, \qquad (\beta=+1, 0, -1)\,.
\end{equation}
%%%%%%%%%%%%%%%
Then, a brane was introduced through the junction conditions and its trajectory and
induced metric were sought for with the condition that an apparent horizon was formed. 
Once again, no black solutions, i.e. solutions with an apparent horizon, with horizon
area larger than that of a black string -- which always arises as a solution of the
field equations -- were found for $A_{AH} \ge {\cal O}(\ell_{AdS}^3)$. As soon as
the horizon radius of the black object fell below the AdS length, black hole solutions
emerged, too.

The analytical studies so far have clearly demonstrated the difficulty in constructing
regular, localised black-hole solutions in the context of a warped spacetime, even in
the simplest of the models. Numerical studies up to now seem to support claims that
the largest class of these objects might not exist at all. Several open questions
remain: why the black string solution was so easy to find but not the same happens
with the black hole one? Is the stability of the sought solutions the answer? 
But the black string solution was also unstable yet it was found -- and its
instability was demonstrated later. If small black holes exist in warped models
why we cannot find not even one by analytical means? And why quantum effects are
so important for large black holes that they forbid their existence as classical
solutions? The community remains divided waiting anxiously for a breakthrough. 

%%%%%%%%%%%%%%%%%%%%%%%%%%%%%%%%%%%%%%%%%%%%%%%%%%%%%%%%%%%%%%%%%%%%%%%%%%%%%%%%

\section{The Flat Case}

From the discussion of the previous section, it becomes clear that, in the context of
the warped models, the only type of black holes that arise naturally from our
field equations are the ones with $r_H \le \ell_{AdS}$. If we are then to study
the class of these small black holes, a particularly convenient limit is the
one where $r_H \ll \ell_{AdS}$. Under this assumption, such a black hole lives,
to first approximation, in a 5-dimensional {\it flat} spacetime with negligible
warping, where the fifth dimension enters the line-element on an equal footing as
the other three spacelike ones. One might wish to generalise the model by adding a
number of additional spacelike dimensions, in which case a contact is made with the
Large Extra Dimensions Scenario: here, a flat higher-dimensional line-element can
be used under the assumption that $r_H \ll {\cal R}$ so that the black hole can not
distinguish the finite-sized extra dimensions from the infinite-sized usual ones. 

The simplest line-element that we may write that describes a spherically-symmetric,
neutral black hole living in a $(4+n)$-dimensional spacetime is the 
Schwarzschild-Tangherlini one \cite{kanTangherlini, kanMP}
%%%%%%%%%%%%%
\begin{equation}
ds^2 = - \left[1-\left(\frac{r_H}{r}\right)^{n+1}\right]\,dt^2 +
\left[1-\left(\frac{r_H}{r}\right)^{n+1}\right]^{-1}\,dr^2 + r^2 d\Omega_{2+n}^2\,,
\label{kanST}
\end{equation}
%%%%%%%%%
where $d\Omega_{2+n}^2$ is the line-element of a $(2+n)$-dimensional unit sphere
%%%%%%%%
\begin{equation}
d\Omega_{2+n}^2=d\theta^2_{n+1} + \sin^2\theta_{n+1} \,\biggl(d\theta_n^2 +
\sin^2\theta_n\,\Bigl(\,... + \sin^2\theta_2\,(d\theta_1^2 + \sin^2 \theta_1
\,d\varphi^2)\,...\,\Bigr)\biggr)\,.
\label{kanunit}
\end{equation}
%%%%%%%%%%%
By applying the Gauss law in $D=4+n$ dimensions, we find for the horizon radius
the result \cite{kanMP}
%%%%%%%%%%%%%
\begin{equation}
r_H = {1\over M_*} \left(M_{BH}\over M_*\right)^{1\over n+1}
\left(8 \Gamma(\frac{n+3}{2}) \over (n+2) \sqrt{\pi}^{(n+1)}\right)^{1/(n+1)}\,.
\label{kanhorizon}
\end{equation}
%%%%%%%%%%%%%%%%
The well-known linear relation between the mass and the horizon radius of the black hole
holding in 4 dimensions arises readily by setting $n=0$ in the above expression, and
substituting the fundamental Planck scale $M_*$ with the 4-dimensional one $M_P$ --
as we will shortly see, the presence of the lower scale $M_*$ in the expression of
$r_H$ plays an important role on deciding whether black holes may be created
at high-energy particle collisions as well as on their properties. 

The line-element (\ref{kanST}) satisfies the higher-dimensional gravitational equations
in the vacuum -- the mass of the black hole is assumed to be much larger than $M_*$
so that the black hole can be considered as a classical object. Since both 
$\Lambda_B$ and the brane self-energy $\sigma$ are assumed to be of the same
order of the fundamental Planck scale, they also may be ignored and as a result no
junction conditions need to be solved, an assumption that simplifies significantly
the study of the black hole. 

The most optimistic, and also the most phenomenologically interesting, scenario is the
one in which $M_*$ is of the order of a few TeV. In that case, collider experiments with
$E>M_*$ can access the so-called trans-planckian energy regime and probe strong gravity
phenomena! The idea that heavy, extended objects, such as p-branes and black holes,
could then produced at ground-based accelerators during high-energy scattering
experiments was put forward in \cite{kanBF}. If, during such a process, the impact
parameter $b$ between the colliding particles is larger than the Schwarzschild radius
$r_H(E)$ that corresponds to the center-of-mass energy of the experiment, elastic and
inelastic processes will take place, dominated by the exchange of gravitons. If, on
the other hand, $b<r_H(E)$, a black hole will be formed according to the Thorne's
Hoop Conjecture \cite{kanThorne} (see also \cite{kanNewHoop} for a higher-dimensional
formulation of the same argument).

Equivalently, a black hole could be created if the Compton wavelength $\lambda_C=4\pi/E$
of the colliding particle of energy $E/2$ lies within the corresponding Schwarzschild
radius $r_H(E)$ \cite{kanMR}. By using the expression for the horizon radius (\ref{kanhorizon}),
the above is written as
%%%%%%%%%%%%%%
\begin{equation}
\frac{4\pi}{E} < {1\over M_*} \left(E\over
M_*\right)^{1\over n+1} \left(8 \Gamma(\frac{n+3}{2}) \over (n+2)
\sqrt{\pi}^{(n+1)}\right)^{1/(n+1)}\,,
\label{kancriterion}
\end{equation}
%%%%%%%%%%%%%%
and leads to the values of $x_{min}=E/M_*$ as a function of the number of extra dimensions
$n$,  necessary for the creation of the black hole, shown in Table \ref{kanxmin}.
From these, we conclude that the center-of-mass energy of the collision must
be approximately one order of magnitude larger than the fundamental Planck
scale $M_*$ -- we remind the reader that the maximum center-of-mass energy
that can be achieved at the Large Hadron Collider at CERN is 14 TeV. 

%%%%%%%%%%%%%%%%%%%%%%%%%%%%%%
%\begin{center}
\begin{table}[t]
\caption{The values of the ratio $x_{min}=E/M_*$, necessary for the creation of 
a black hole, as a function of $n$.}
\begin{center}
\begin{tabular}{cccccc}  \br %\hline\noalign{\smallskip}
$n=2$\, & \,$n=3$ \,& \,$n=4$\, & \,$n=5$ \,& \,$n=6$\, & \,$n=7$ \\ \mr
%\noalign{\smallskip}\noalign{\smallskip} \mr
$x_{min}=8.0$ & $x_{min}=9.5$ & $x_{min}=10.4$ & $x_{min}=10.9$
& $x_{min}=11.1$ & $x_{min}=11.2$\\ \br
\end{tabular} 
\end{center}
\label{kanxmin}
\end{table}
%\end{center}
%%%%%%%%%%%%%%%%%%%%%%%%%%%%%%

We should note here that only a Quantum Theory of Gravity could offer a detailed
analysis of a trans-planckian collision between two particles leading to the creation of
a black hole. As the complete, consistent form of such a theory is still missing,
this problem is most often modelled by the collision of two gravitational Aichelburg-Sexl
shock waves \cite{kanAS} moving at the speed of light. Solving the boundary value
problem of the creation or not of a closed trapped surface, the maximum value of the
colliding impact parameter $b_{\rm max}$ and the mass of the produced black hole can be found
\cite{kanPenrose, kanDP, kanCardoso1, kanCardoso3, kanEG, kanYN, kanYR, kanYoshMann}. The value
of the former parameter can also lead to an estimate of the production cross-section
of the black holes through the relation $\sigma_{\rm production} \simeq \pi b^2_{\rm max}$,
valid for a high-energy collision. The most recent analyses seem to agree quite well
with some early estimates \cite{kanGiddings, kanGT, kanDL} according to which the production
cross-section takes the values $10^5\,{\rm fbar}$ and $10^1\,{\rm fbar}$ for a black
hole with mass $M_{BH}=5\,{\rm TeV}$ and $M_{BH}=10\,{\rm TeV}$, respectively, and
under the assumption that $M_*=1\,{\rm TeV}$ and $D=10$. In both cases, the value
of the production cross-section is considered to be significant for beyond the Standard
Model processes.

Let us now briefly discuss the properties of the produced black holes \cite{kanreview,
kanADMR}. The value of the horizon radius as a function of $n$ may easily follow
from Eq. (\ref{kanhorizon}) upon particular choices for $M_*$ and $M_{BH}$. 
Assuming that $M_*=1$ TeV and $M_{BH}=5$ TeV, the corresponding values of $r_H$
are shown in Table \ref{kanprop}. In the same Table, the values of the black hole
temperature, that for the simple black-hole prototype (\ref{kanST}) is given by the equation
%%%%%%%%%%%%%
\begin{equation}
T_{H} = \frac{k}{2\pi}=
\frac{1}{4\pi}\,\frac{1}{\sqrt{|g_{tt}\,g_{rr}|}}\left(\frac{d|g_{tt}|}{dr}
\right)_{r=r_H}={(n+1) \over 4\pi\,r_H}\,,
\label{kantemp}
\end{equation}
%%%%%%%%%%%%%
are also shown. Note that the values of $r_H$ lie in the subnuclear regime while those
of $T_H$ in the GeV range - both regimes are easily accessible by present day
experiments and detection techniques.

%%%%%%%%%%%%%%%%%%%%%%%%%%%%%%%%%%%%%%%%%%%%%%%%%%
\begin{table}[b]
\caption{Horizon radius and temperature of the Schwarzschild-Tangherlini black 
hole as a function of the number of extra dimensions, for $M_*=1$\,TeV and
$M_{BH}=5$\,TeV  \cite{kanreview}}
\begin{center}
\begin{tabular}{cccccccc} \br %\hline\noalign{\smallskip}
\hspace*{0.5cm} $n$ \hspace*{0.5cm}& \hspace*{0.5cm} 1 \hspace*{0.5cm} & \hspace*{0.5cm} 2 
\hspace*{0.5cm} & \hspace*{0.5cm} 3 \hspace*{0.5cm} & \hspace*{0.5cm} 4 \hspace*{0.4cm}
& \hspace*{0.5cm} 5 \hspace*{0.5cm} & \hspace*{0.5cm} 6 \hspace*{0.5cm} &
\hspace*{0.5cm} 7 \hspace*{0.5cm} \\ \mr 
%\noalign{\smallskip}\svhline\noalign{\smallskip}
$r_H$ ($10^{-4}$ fm) \hspace*{0.0cm} & 4.06 & 2.63 & 2.22 
& 2.07 & 2.00 & 1.99 & 1.99 \\[1mm]
$T_{H}$ (GeV) & 77 & 179 & 282 & 379 & 470 
& 553 & 629 \\ \br
\end{tabular}
\end{center}
\label{kanprop}
\end{table}
%%%%%%%%%%%%%%%%%%%%%%%%%%%%%%%%%%%%%%%%%%%%%%%%%

The non-vanishing temperature gives rise to a thermal type of radiation emitted by
the black hole in the form of elementary particles. This is the well-known Hawking
radiation whose energy spectrum is given by the expression \cite{kanHawking, kanUnruh}
%%%%%%%%%%%%%%%%
\begin{equation}
\frac{\textstyle dE(\omega)}
{\textstyle dt} = \int\,\frac{\textstyle {  |{\cal A}(\omega)|^2} \,\,\omega}
{\textstyle \exp\left(\omega/T_{H}\right) \mp 1}\,\,
\frac{\textstyle d\omega}{\textstyle (2\pi)}\,.
\label{kanrateS}
\end{equation}
%%%%%%%%%%
In the above, $\omega$ is the energy of the emitted particles and $\pm 1$ a statistics
factor for fermions and bosons, respectively. The origin of the Hawking radiation
is quantum-mechanical since classically nothing is allowed to escape through the
horizon of the black hole: a virtual pair of particles is produced outside the horizon
of the black hole; when the antiparticle falls inside the black hole and the particle
is allowed to propagate towards infinity, the black hole emits radiation characterized
by its temperature. Its spectrum is of a black-body type except for the factor 
${|{\cal A}(\omega)|^2}$ appearing in the numerator of the emission rate, called the
Absorption  Probability (or, {\it greybody factor}). Its presence is due to the fact
that not all of the outward propagating particles will reach infinity - some of them,
will be reflected back due to the strong gravitational field of the black hole.  

The emission of Hawking radiation causes the black hole to evaporate and thus to
have a finite lifetime. For the same values of $M_*$ and $M_{BH}$ as above, the typical
lifetime of the black hole comes out to be $\tau=(1.7-0.5) \times 10^{-26}$ sec for
$n=1-7$. Contrary to the large astrophysical black holes, whose lifetime exceeds that
of the universe, these microscopic black holes will evaporate almost instantly after
their creation. Another important factor that needs to be taken into account, as it
also determines the rate of evaporation, is that a higher-dimensional black hole
emits simultaneously radiation in two channels, the `brane' and the `bulk' one.
The first one involves all the Standard Model Particles living in our 4-dimensional world,
fermions, gauge bosons and Higgs-like scalars; the second emits mainly gravitons
and possibly other particles, like scalar fields, that are allowed to live in
the bulk provided that they carry no quantum charges under the Standard Model
gauge group. Although the `brane' channel is the most phenomenologically interesting
to us as it can be directly measured by a brane observer, the `bulk' one is almost
of equal importance as we need to know how much black-hole energy is lost in the 
non-accessible transverse dimensions.  

In what follows, we will therefore start with the study of the emission of Hawking
radiation on the brane, and then we will discuss what is known for the emission in
the bulk. We will consider the emission during the two intermediate stages of the
life of a black hole, the axially-symmetric {\it spin-down} phase and the
spherically-symmetric {\it Schwarzschild} one. These two phases are preceded
by the {\it balding} phase, in which the black hole first forms and settles down
to a state that satisfies the no-hair theorem of General Relativity, and are
followed by the {\it Planck} phase, in which the black hole turns to a quantum
object with unknown properties.

Let us start with the Hawking radiation emitting phase with the simplest gravitational
background, the Schwarzschild one. Its line-element is given by Eq. (\ref{kanST})
and it is the background seen by any particle propagating in the higher-dimensional
spacetime. But for a particle localised on the brane, like the Standard Model fields,
we need to consider the projection of (\ref{kanST}) on the brane - this is done by
setting the values of all additional $\theta_i$ coordinates,
with $i=2,...,n+1$, introduced to describe the additional spacelike dimensions,
to $\frac{\pi}{2}$. Then, the resulting brane line-element assumes the form
%%%%%%%
\begin{equation}
ds^2_4 = - \left[1-\left(\frac{r_H}{r}\right)^{n+1}\right] dt^2 +
\left[1-\left(\frac{r_H}{r}\right)^{n+1}\right]^{-1} dr^2 + 
r^2\,d\Omega_2^2\,.
\label{kanprojectedSchw}
\end{equation}
%%%%%%%%%%%%%%%
The non-trivial $n$-dependence of the projected-on-the-brane background will introduce
a similar dependence in the field equations. By using the Newman-Penrose method
\cite{kanNP, kanChandra}, these take the form of a Teukolsky-like \cite{kanTeukolsky}
``master'' equation of motion with the spin $s$ appearing as a parameter \cite{kanreview,
kanKMR1, kanKMR2}. By using also a factorized ansatz for the wavefunction of the field
of the form 
%%%%%%%%%%%%%%%%%%%
\begin{equation}
\Psi_s=e^{-i\omega t}\,e^{im\varphi}\,\Delta^{-s}\,R_s(r)\,S_{s l}^m(\theta)\,,
\end{equation}
%%%%%%%%%%%%%%
the general field equation reduces to two decoupled equations, one for the radial function
$R_s(r)$ and one for the spin-weighted spherical harmonics \cite{kanGoldberg}
$S_{s l}^m(\theta)$
%%%%%%%%%%%%%%%
\begin{equation}
\Delta^s \frac{\textstyle d}
{\textstyle dr}\left(\Delta^{1-s}\frac{\textstyle d R_s^{~}}
{\textstyle dr}\right)+ \left[\frac{\textstyle \omega^2 r^2}{\textstyle h}+2i\omega s r-
\frac{\textstyle is\omega r^2 h'}{\textstyle h}
-\lambda_{s l}\right]R_s (r)=0\,,
\end{equation}
%%%%%%%%%%%%
and
%%%%%%%%%%%%%
\begin{equation}
\frac{1}{\sin\theta}\,\frac{\textstyle d}
{\textstyle d\theta}\left(\sin\theta\frac{\textstyle d S^m_{s l}}
{\textstyle d\theta}\right)+ \left[-\frac{\textstyle 2ms\cot\theta}{\textstyle \sin\theta}
-\frac{m^2}{\sin^2\theta}+s-s^2\cot^2\theta+\lambda_{s l}\right]S^m_{s l}(\theta)=0\,,
\end{equation}
%%%%%%%%%%%%%%
respectively. In the above, we have defined the function $\Delta\equiv r^2\,h
\equiv r^2\left[1- \left(\frac{r_H}{r}\right)^{n+1}\right]$, while
$\lambda_{s l}=l(l+1)-s(s-1)$ is the eigenvalue of the spin-weighted
spherical harmonics. By setting the values $s=0,1/2$ and 1, one may easily derive from
the above the corresponding radial and angular equations for brane-localised scalars,
fermions and gauge bosons.

%%%%%%%%%%%%%%%%%%%%
\begin{figure}[b]
\begin{center}
\includegraphics[width=7cm, height=5cm]{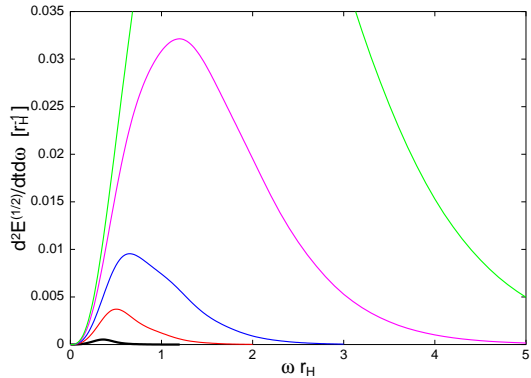}
\end{center}
\caption{\label{kanfermions}Energy emission rates for fermions on the
brane or $n=0,1,2,4$ and 6 (from bottom to top) \cite{kanreview}.}
\end{figure}
%%%%%%%%%%%%%%%%%%%%%%

The angular equation carries no useful information as, given the symmetry of the
gravitational background, the emitted radiation is evenly distributed over a $4\pi$
solid angle. But the radial equation needs to be explicitly solved as the solution
for the radial function $R_s(r)$ determines the Absorption Probability through the
formula
%%%%%%%%%%%%%%
\begin{equation}
|{\cal A}(\omega)|^2 \equiv 1-|{\cal R}(\omega)|^2 \equiv \frac{\textstyle 
\hspace*{2mm}{\cal F}_{\rm horizon_{~}}} 
{\textstyle {\cal F}^{~}_{\rm infinity}}\,,
\end{equation}
%%%%%%%%%%%%%%
where ${\cal R}(\omega)$ is the Reflection coefficient and ${\cal F}$ the flux
of energy towards the black hole. The Absorption Probability can be computed
both analytically \cite{kanKMR1, kanKMR2, kanIOP1} and numerically \cite{kanHK}.
Whereas the first method can offer a more qualitative insight into the details
of the emission of different species of fields, the second provides us with exact
quantitative results for the greybody factor. When this is combined with the
expression for the temperature of the black hole in the context of Eq. (\ref{kanrateS}),
we obtain the radiation spectra for scalars, fermions, and gauge bosons. The
differential emission rate per unit time and frequency, in terms of the number of
transverse spacelike dimensions $n$, for the indicative case of fermions, is
shown in Fig. \ref{kanfermions}. What we may easily observe -- with the same
holding for scalars and gauge bosons -- is that the energy emission rate on the
brane is greatly enhanced by the number of additional spacelike dimensions. 
The same effect is shown in numbers in the entries of Table \ref{kantableS},
where the total emissivities have been normalised to the ones for $D=4$: as
the value of $n$ increases the enhancement factors reach 3 or 4 orders of
magnitude  \cite{kanHK}.

\medskip
%%%%%%%%%%%%%%%%%
\begin{table}[t]
\caption{Total emissivities for brane-localised scalars, fermions and gauge
bosons \cite{kanHK}}
\begin{center}
\begin{tabular}{ccccccccc} \br
$n$ & 0  & 1 & 2 & 3 & 4 & 5 & 6 & 7 \\ \mr  
 Scalars  & 1.0 & 8.94 & 36.0 & 99.8 & 222 &
429 & 749 & 1220\\ 
 Fermions  & 1.0 & 14.2 & 59.5 & 162 & 352 &
664 & 1140 & 1830\\ 
G. Bosons  & 1.0 & 27.1 & 144 & 441 & 1020 &
2000 & 3530 & 5740 \\ \br
\end{tabular}
\label{kantableS}
\end{center}
\end{table}
%%%%%%%%%%%%%%%%%%

\begin{figure}[b]
\begin{center}
\hspace*{0.0cm}\hbox{
\scalebox{0.55}{\rotatebox{0} 
{\includegraphics[width=12cm, height=9.5cm]{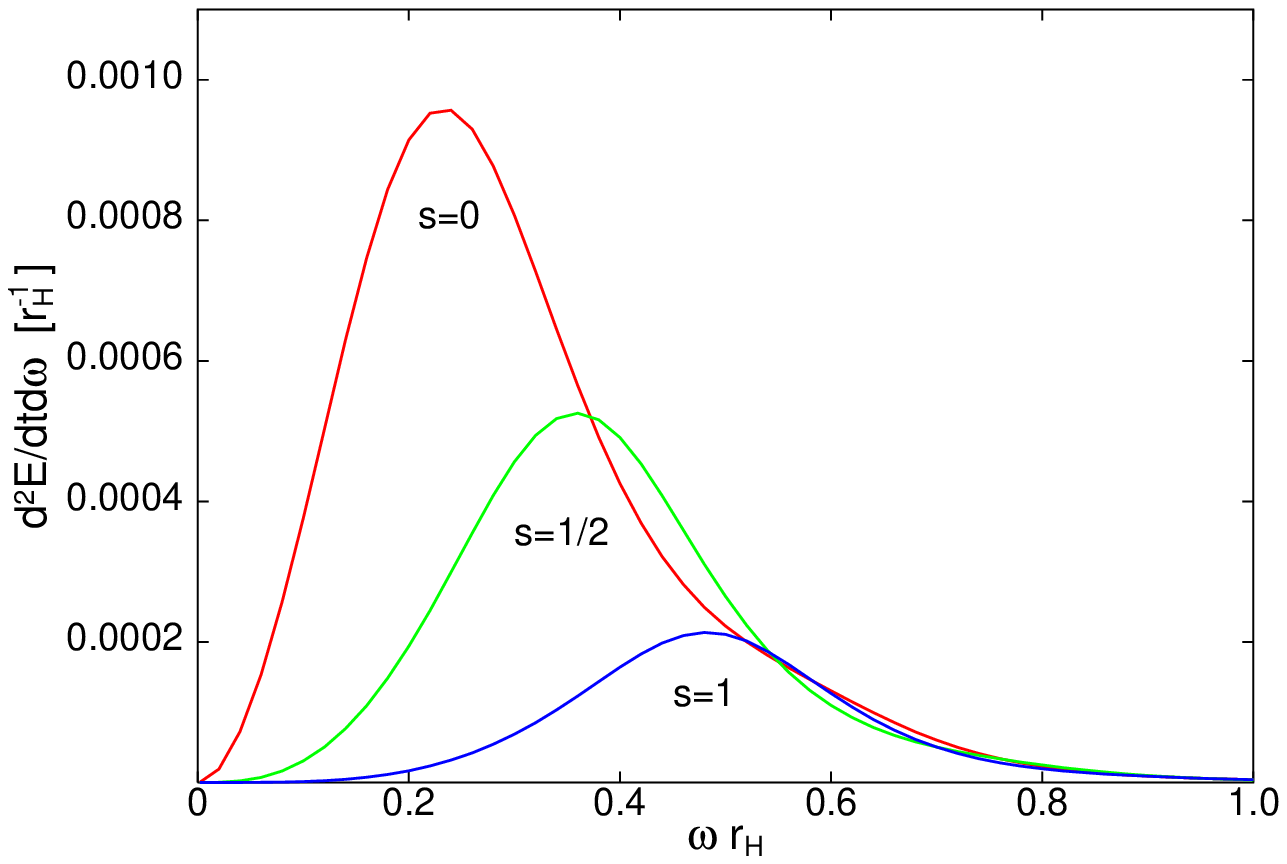}}}
\hspace*{0.6cm}\scalebox{0.55}
{\rotatebox{0}{\includegraphics[width=12cm, height=9.5cm]{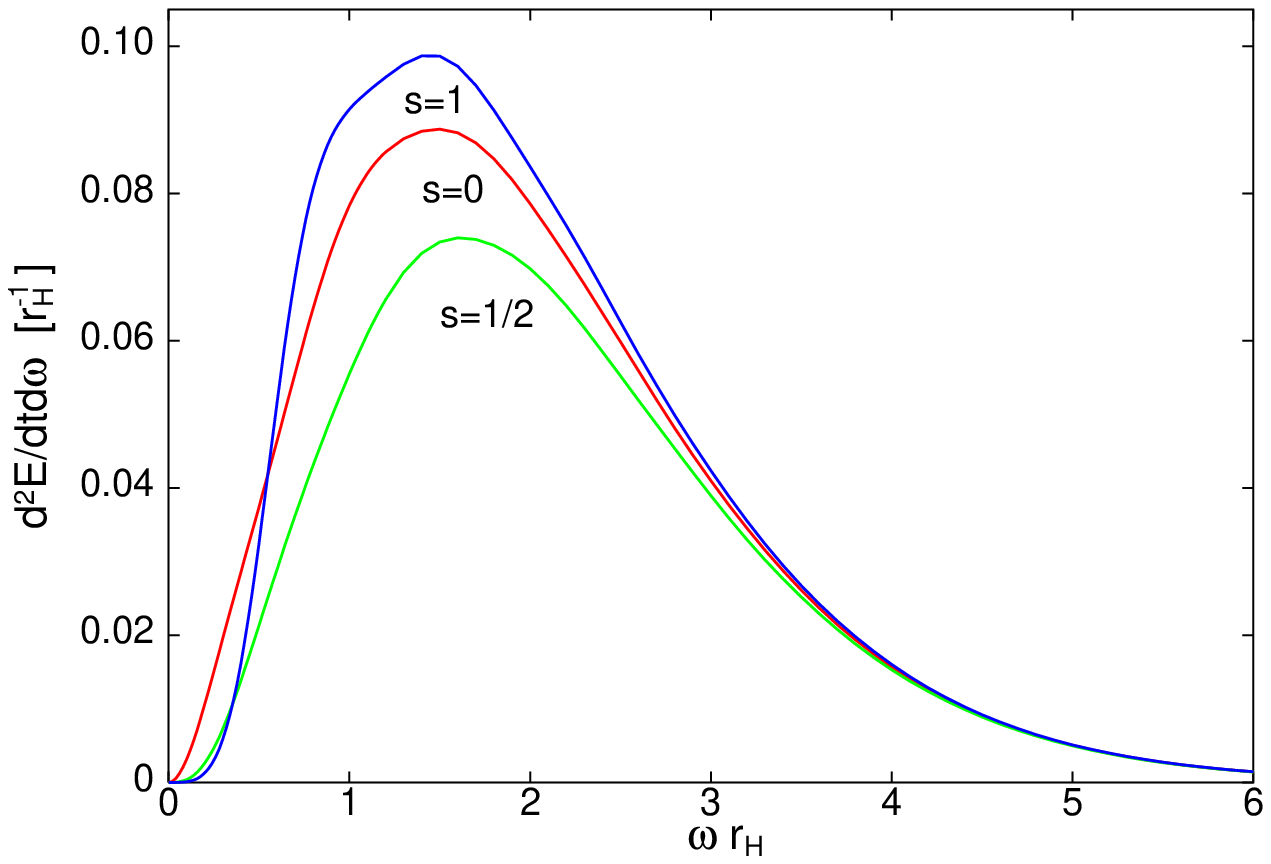}}}}
\caption{Energy emission rates for brane fields for $n=0$ (left
plot) and $n=6$ (right plot) \cite{kanHK}.}
\label{kanrelative}
\end{center}
\end{figure}

In the context of the same work, it was also shown
that not only the emission rate but also the type of radiation emitted by the black
hole depends strongly on $n$. In Fig. \ref{kanrelative}, the energy emission rates,
for scalars, fermions and gauge bosons, are depicted for the two indicative cases of
$n=0$ and $n=6$ \cite{kanHK}. It is clear that, whereas a 4-dimensional spherically-symmetric
black hole prefers to emit scalar fields, a 10-dimensional one exhibits a clear preference
for gauge bosons. This feature may also be used as an observable signature that will
help us determine the number of additional dimensions in nature through the observation
of the emitted Hawking radiation from a microscopic black hole.  

%%%%%%%%%%%%%%%%%%%%
%%%%%%%%%%%%%%%%%%%%%%%%

The gravitational background around a higher-dimensional, rotating black hole is
significantly more complicated and takes the form of the Myers-Perry solution \cite{kanMP}
%%%%%%%%%%
\begin{eqnarray}
\hspace*{-1cm}
ds^2 &=& \left(1-\frac{\mu}{\Sigma\,r^{n-1}}\right)dt^2+\frac{2 a\mu\sin^2\theta}
{\Sigma\,r^{n-1}}\,dt\,d\varphi-\frac{\Sigma}{\Delta}dr^2 \nonumber \\[1mm] 
&-& \hspace*{0.3cm}
\Sigma\,d\theta^2-\left(r^2+a^2+\frac{a^2\mu\sin^2\theta}{\Sigma\,r^{n-1}}\right)
\sin^2\theta\,d\varphi^2 - r^2 \cos^2\theta\,d\Omega^2_n\,,
\label{kanMPmetric}
\end{eqnarray}
%%%%%%%%
where
%%%%%%%%%
\begin{equation}
\Delta=r^2+a^2-\frac{\mu}{r^{n-1}}\,, \quad \quad\Sigma=r^2+a^2\cos^2\theta\,.
\end{equation}
%%%%%%%%%%%
In the above, we have chosen a simple case of the Myers-Perry class of rotating
solutions, namely the one where the black hole has only one non-vanishing angular
momentum component -- this is due to the fact that the black hole has been
created by particles that are localised on the brane and have non-zero impact
parameters only along a brane spacelike coordinate. The parameters $\mu$ and $a$
are then associated to the black hole mass and angular momentum, respectively,
through the relations
%%%%
\begin{equation}
M_{BH}=\frac{(n+2) A_{2+n}}{16 \pi G}\,\mu \qquad {\rm and} \qquad 
J=\frac{2}{n+2}\,a\,M_{BH}\,, \label{kan-mu}
\end{equation}
%%%%%%%%%%%%%%%%%
where $A_{2+n}$ is the area of a $(2+n)$-dimensional unit sphere. The horizon
radius is found by setting $\Delta(r_H)=0$ and is found to be:
$r_H^{n+1}=\mu/(1+a_*^2)$, where we have defined the quantity $a_* \equiv a/r_H$.
Finally, the temperature and rotation velocity of this black hole are given by
%%%%%%%%%%%%
\begin{equation}
T_{H}=\frac{(n+1)+(n-1)\,a_*^2}{4\pi(1+a_*^2)\,r_{H}}\,,
\qquad \Omega_{H}=\frac{a}{(r_H^2+a^2)}\,. \label{kan-Temprot}
\end{equation}

The derivation of the radiation spectra in the case of the spin-down phase follows
along the same lines as in the case of the Schwarzschild phase. We first need to
find the line-element seen by the brane-localised Standard Model fields - this
follows again by fixing the values of the ``extra'' angular coordinates in which
case the $d\Omega_{n}^{2}$ part of the metric (\ref{kanMPmetric}) disappears leaving
the remaining unaltered. By employing again the Newman-Penrose method and a similar
factorised ansatz for the spin-$s$ field perturbation -- with the spin-weighted spherical
harmonics being now replaced by the spin-weighted spheroidal harmonics \cite{kanFlammer},
we derive two decoupled master equations, one for the radial part of the field and
one for the angular part, namely \cite{kanreview, kanCKW}
%%%%%%%%%%
\begin{equation}
\Delta ^{-s}\,\frac {d}{dr} \left(\Delta^{s+1}\,
\frac {d R_s}{dr}\right) + \left[\frac{K^2-iKs\Delta '}{\Delta}+4i s\omega r +
s\left( \Delta '' -2 \right)\delta_{s,|s|}-\Lambda^m_{sj}\right] R_s=0 
\label{kan-radialrot}
\end{equation}
%%%%%%%%%%%%%%%%
and
%%%%%%%%%%%%%%%%
\begin{eqnarray} && \hspace*{-1.0cm}\frac{1}{\sin\theta}\,\frac{\textstyle d}
{\textstyle d\theta}\left(\sin\theta\frac{\textstyle d S^m_{sj}}
{\textstyle d\theta}\right)+ \left[-\frac{\textstyle 2ms\cot\theta}{\textstyle \sin\theta}
-\frac{m^2}{\sin^2\theta}+a^2\omega^2\cos^2\theta\right. \nonumber \\[4mm]
&&\hspace*{3cm} \left.- 2 a \omega s \cos\theta +
s-s^2\cot^2\theta+\lambda_{sj}\right]S^m_{sj}(\theta)=0\,,
\label{kan-angulareq}
\end{eqnarray}
%%%%%%%%%%%%%%%%%
where we have used the definitions
%%%%%%%%%%%%%%
\begin{equation}
K=(r^2+a^2)\,\omega-am\,, \qquad \Lambda^m_{sj}=\lambda_{sj} + a^2 \omega^2
-2 am \omega\,.
\end{equation}
%%%%%%%%%%%%%%%%

\begin{figure}[t]
\begin{center}
\mbox{ 
\includegraphics[scale=0.6]{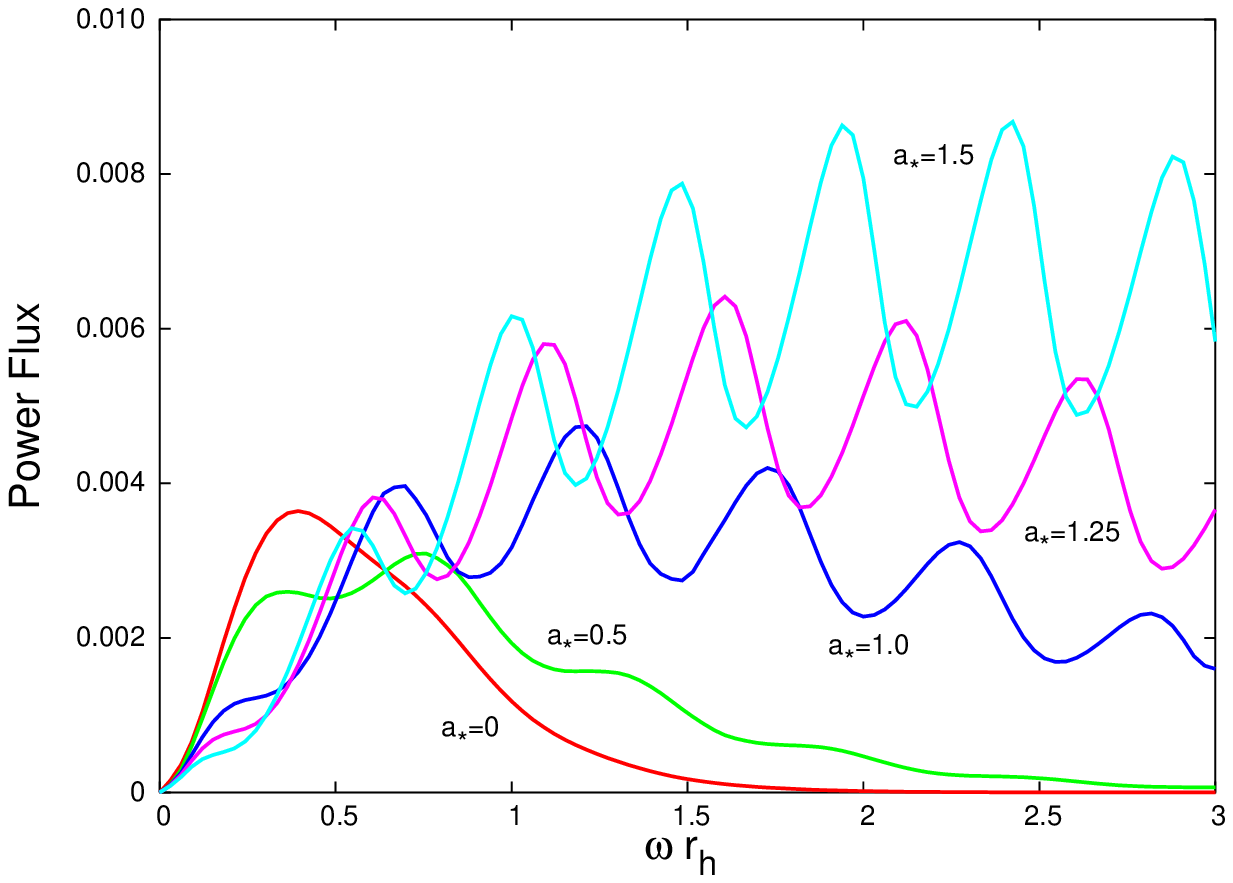}}
{\includegraphics[scale=0.6]{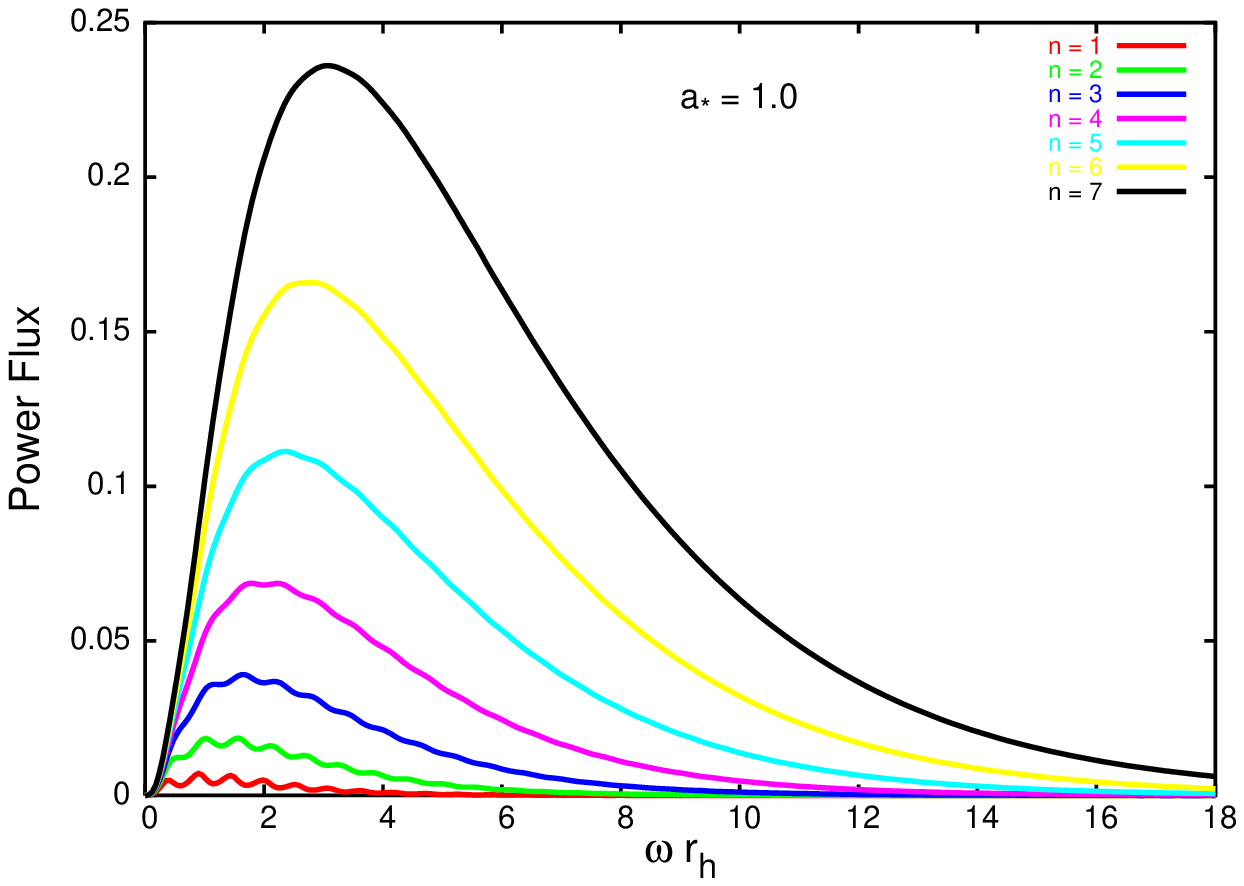}}
\caption{\label{kanratesrotplot} Energy emission rates for brane-localised scalar
fields in terms of the angular parameter (left plot) \cite{kanDHKW} and gauge bosons
in terms of the number of extra dimensions (right plot) \cite{kanCKW}.}
\end{center}
\end{figure}

The angular eigenvalue $\lambda_{sj}$ connecting the radial and angular equations does
not exist in closed form. It may be computed either analytically, through a power series
expansion in terms $a \omega$ \cite{kanStaro, kanFackerell, kanSeidel} or numerically
\cite{kanCKW, kanHK2, kanDHKW, kanCDKW}. In the same way, the Absorption Probability
can be found by solving the radial equation, for all species of fields, either
analytically \cite{kanCEKT2, kanCEKT3} or numerically \cite{kanCKW, kanHK2, kanDHKW,
kanCDKW, kanFS, kanIOP2, kanIOP3}. As before, the exact spectra follow only by means of
numerical analysis by using the derived value of the Absorption Probability together
with the ones of $T_H$ and $\Omega_H$ in the context of the formula
\cite{kanHawking, kanUnruh, kanOW}
%%%%%%%%%%%%%%
\begin{equation}
\frac{d^2\,E}{\,dt\,d\omega}= \frac{1}{2\pi}\sum_{j,m}\,
\frac{{  |{\cal A}(\omega)|^2}\,\omega}{\exp(\tilde\omega/T_H)\mp 1}\,, 
\end{equation}
%%%%%%%%%
where $\tilde \omega= \omega -m \Omega_H$. The radiation spectra will now depend on
two topological parameters, the angular-momentum parameter $a_*$ of the black hole
and the number of additional spacelike dimensions $n$. In Fig. \ref{kanratesrotplot},
we present the energy emission rates, for the indicative cases of brane-localised
scalars and gauge bosons, in terms of these two parameters. We may easily conclude
that the energy emission rates are significantly enhanced by both parameters: the
enhancement factor is of order ${\cal O}(10)$ in terms of $a_*$ and of order
${\cal O}(100)$ in terms of $n$. As $n$ increases, the gauge bosons are again the
species of particles that a rotating black hole increasingly prefers to emit. 

%%%%%%%%%%%%%%%%%%%%%%%%%%%%%%%%%%%%%%%%%%%%%%%%%%%%
%%%%%%%%%%%%%%%%%%%%%%%%%%%%%%%%%%%%%%%%%%%%%%%%%%%%

Finally, let us briefly add that, contrary to the case of a spherically-symmetric
black hole, the emission of Hawking radiation by a rotating black hole has a 
non-trivial angular distribution in space. As it was shown in detail in \cite{kanCKW, 
kanDHKW, kanCDKW} by numerically solving the angular equation (\ref{kan-angulareq}),
the centrifugal force causes all particles with intermediate
and high frequency to be emitted along the equatorial plane, i.e. vertically to
the rotation axis of the black hole. In addition, the emission of particles with
non-vanishing spin and low frequency is polarised along the rotation axis with the
effect being more prominent the larger the value of $s$ is. 

Let us now focus on the emission of particles in the bulk by the higher-dimensional
black hole. Although the bulk emission can never be detected by a brane observer,
it determines the amount of energy left for emission on the brane. The black hole
can emit only particular species of particles in the bulk,
namely gravitons and possibly scalar fields. We will first consider the emission of
the latter species during the Schwarzschild phase. By following a similar analysis
as the one for emission on the brane, the bulk scalar equation can be solved in the
background of Eq. (\ref{kanST}) and the corresponding greybody factor can be found
both analytically \cite{kanKMR1, kanFS0} and numerically \cite{kanHK}. The exact
numerical results allow us to compute the complete radiation spectra that exhibit
a similar behaviour to the ones for brane emission, namely a strong enhancement 
with the number of additional spacelike dimensions. More importantly, they allow
us to compute the relative emissivity for scalar fields, i.e. the ratio of the 
energy emitted by the black hole per unit time over the whole frequency regime 
in the two emission channels, `bulk' and `brane'. The values of this ratio as
a function of $n$ is shown in Table \ref{kan-ratioS} \cite{kanHK}. From these, it becomes  
clear that the brane scalar channel remains the dominant one for all values of
$n$, however, as $n$ takes large values, the emission in the two channels becomes
comparable.

%%%%%%%%%%%%%%%%%%%
\begin{table}[t]
\caption{Bulk-to-Brane Relative Emissivities Ratio for scalar fields in terms of $n$
\cite{kanHK}}
\begin{center}
\begin{tabular}{ccccccccc}  \br 
$n$  & \hspace*{0.4cm} $0$ \hspace*{0.2cm} & \hspace*{0.2cm} $1$ 
\hspace*{0.1cm} & \hspace*{0.1cm} $2$ \hspace*{0.1cm} & \hspace*{0.1cm} $3$
\hspace*{0.1cm} & \hspace*{0.1cm} $4$ \hspace*{0.1cm} & \hspace*{0.1cm} $5$ 
\hspace*{0.1cm} & \hspace*{0.1cm} $6$ \hspace*{0.1cm} & \hspace*{0.1cm} $7$
\hspace*{0.1cm} \\ \mr
\hspace*{0.1cm} {\rm Bulk/Brane}
\hspace*{0.3cm}
& 1.0 & 0.40 & 0.24 & 0.22 & 0.24 & 0.33 & 0.52 & 0.93\\ \br
\end{tabular}
\end{center}
\label{kan-ratioS}
\end{table}
%%%%%%%%%%%%%%%%%%

Similar analyses, analytical \cite{kanCNS, kanCEKT1} as well as numerical \cite{kanCCG,
kanJP}, have been performed for the emission of gravitons in the bulk during the
Schwarzschild phase. The results produced in \cite{kanCCG} have revealed an enhancement
factor for the energy emission rate for gravitons with $n$ that surpasses the one for
any other species of field, reaching the value of $10^{5-6}$. Nevertheless, contrary 
to our naive expectation this is not enough to render the bulk channel the dominant
one: the total number of brane degrees of freedom is larger than the one for 
gravitons living in the bulk, and overall the brane channel still dominates over
the bulk one, thus offering strong support to an early work \cite{kanEHM} where a
similar claim was made\footnote{For some special cases where the bulk channel may
dominate over the brane one, see \cite{kanBGK, kanCCDN}.}.

%%%%%%%%%%%%%%%%%%%%%%%%%%%%%%%%%%%%%%%%%%%%%%%%%%%%
\begin{figure}[b]
\begin{center} 
\includegraphics[scale=0.8]{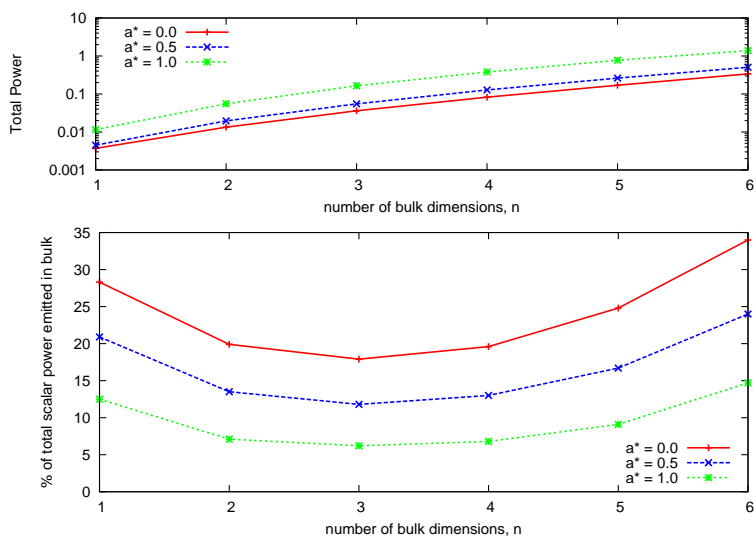}
\caption{\label{kans0abulk} Total power emitted by a rotating black hole in the scalar
channel (upper plot) and the $\%$ of this power emitted in the bulk (lower plot)
\cite{kanCDKW2}.}
\end{center}
\end{figure}
%%%%%%%%%%%%%%%%%%%%%%%%%%%%%%%%%%%%%%%%%%%%%%%%%%%%

The study of the emission of fields in the bulk during the spin-down phase has
not been completed yet. Until now only the emission of scalar fields has been
studied \cite{kanJP05, kanCEKT4, kanCDKW2} as the field equations for gravitons
in an axially-symmetric gravitational background remains largely unknown. In
\cite{kanCDKW2}, a comprehensive study of the scalar emission in the bulk was
made and compared to the one on the brane. It was demonstrated that although 
the total energy output of the black hole in the scalar channel increases with
the angular momentum of the black hole, the bulk emission is actually suppressed
especially in the low $a_*$ regime (see Fig. \ref{kans0abulk}).
This is due to the fact that the temperature
of the black hole decreases with $a_*$, so it is left to the behaviour of the
greybody factor to determine the final form of the spectrum: whereas
$|{\cal A}(\omega)|^2$ for brane scalars increases significantly with $a_*$
overcoming the decrease in temperature, for bulk scalars its enhancement is not
large enough to achieve the same effect. As a result, the brane dominance 
persists even in the spin-down phase, at least in the scalar channel of emission.

\section{Conclusions}

In this review talk, we have discussed some recent as well as some less recent
attempts to construct and study black-hole solutions in the context of theories
postulating the existence of additional spacelike dimensions in nature. These 
attempts have been proven to be very challenging in some cases and very informative
in others.

In the context of theories with Warped Extra Dimensions, all efforts to construct
analytically a regular black hole solution have so far failed. Different approaches
have been developed over the years, where either the bulk or the brane solution
played the central role, however none of them managed to achieve a breakthrough.
The long-sought regular, localised, static black hole remains elusive whereas
its `twin' solution, the black string, readily emerges although unstable.
Theoretical arguments and numerical studies seem even to support that the
existence of large, static black holes is highly disfavoured in warped models. 

Whereas constructing a regular black hole in a spacetime filled with a negative
cosmological constant is highly non-trivial, in higher-dimensional flat spacetime
this task is straightforward with regular black-hole solutions being known for
decades now. The theoretical challenges of the existence of these black holes are,
in the context of the Large Extra Dimensions model, removed -- the interest is 
turned instead to the study of their properties and any phenomenological implications
that the potential creation of very small black holes at high-energy particle
collisions may have. If strong gravity emerges indeed at energies close to 1 TeV, 
black holes may even be produced at ground-based colliders such as the Large Hadron
Collider (LHC). One of the most distinctive signatures should be the Hawking radiation
spectra from which a wealth of information can be derived on the emitted particle
properties and the topological parameters of our spacetime.

The frontiers in both areas remain open -- the outcome of our efforts to construct
regular black hole solutions in the context of warped models will show how realistic
these models, inspired from superstring theory, really are. On the other hand, the
potential detection of a microscopic black hole, guided by our predictions for their
properties and observable signatures, will prove the validity of the theoretical
foundation of black hole physics we have built both in 4-dimensional and 
higher-dimensional spacetime.

%%%%%%%%%%%%%%%%%%%%%%%%%%%%%%%%%%%%%%%%%%%

\ack I am grateful to my collaborators (K. Tamvakis, I. Olasagasti, J. March-Russell,
C. Harris, A. Barrau, J. Grain, G. Duffy, E. Winstanley, M. Casals, S. Creek, 
O. Efthimiou, R. Gregory, B. Mistry and S. Dolan, in chronological order) for
our enjoyable and fruitful collaborations. I would also like to thank the organisers
of the ``NEB XIII -- New Developments in Gravity'' conference for their kind invitation
to give this review talk. I finally acknowledge partici\-pa\-tion in the RTN Universenet
(MRTN-CT-2006035863-1 and MRTN-CT-2004-503369).

\end{document}